\documentclass[12pt]{article} 
\usepackage{amssymb}
\usepackage{graphicx}

\begin{document} 
\begin{center}
{\large \bf Asymmetric vector mesons produced in nuclear collisions}

\vspace{0.5cm}                   

{\bf I.M. Dremin, V.A. Nechitailo}

\vspace{0.5cm}                       

         Lebedev Physical Institute, Moscow 119991, Russia\\
\medskip

        National Research Nuclear University "MEPhI", Moscow 115409, Russia     

\end{center}

\begin{abstract}

It is argued that the experimentally observed phenomenon of asymmetric shapes
of vector mesons produced in nuclear media during high energy nucleus-nucleus 
collisions can be explained as Fano-Feshbach resonances. It has been observed 
that the mass distributions of lepton pairs created at meson decays decline 
from the traditional Breit-Wigner shape with some excess in the low-mass 
wing of the resonance. It is clear that the whole phenomenon is related 
to some interaction with the nuclear medium. Moreover, it can be further 
detalized in quantum mechanics as the interference of direct and continuum 
states in Fano-Feshbach effect. To reveal the nature of the interaction it is 
proposed to use a phenomenological model of the additional contribution due 
to Cherenkov gluons. They can be created because of the excess of the 
refractivity index over 1 just in the low-mass wing as required by the 
classical Cherenkov treatment. In quantum mechanics, this requirement is 
related to the positive real part of the interaction amplitude in this wing. 
The corresponding parameters are found from the comparison with $\rho $-meson 
data and admit reasonable explanation.
\end{abstract}

\section{Introduction}

Resonance peaks are observed in many natural phenomena. They are treated in
numerous textbooks (see, e.g., \cite{fls}). The traditional way
of their description is to compare their shapes with the relativistic
Breit-Wigner formula \cite{bw, china}
\begin{equation}
f=\frac {k}{(m_{r }^2-M^2)^2+M^2\Gamma _r^2},
\label{bw}
\end{equation}
where $f$ denotes a signal strength, $k$ is the normalization constant, 
$M$ is the scanning energy, $m_r$ is the maximum position (the resonance
mass), $\Gamma _r$ is the resonance width. One measures the signal intensity
$f$ at different energies $M$. For example, the atoms considered as oscillators emit 
as Breit-Wigner resonances. The resonance shape is a purely statistical
phenomenon. It depends on many details of interactions and need not to be
necessarily symmetric.

The asymmetric resonance peaks were experimentally observed in various
fields of physics even before the formula (\ref{bw}) was proposed. 
E.g., the resonance of He atoms observed in the inelastic scattering of
electrons is strongly asymmetric (see \cite{rice, fano}). Many spectroscopic
studies of atoms were devoted to this phenomenon. The name of Fano
resonances was attached to observed asymmetric resonances. In nuclear physics, 
they are known as Feshbach resonances \cite{fesh}.

Both types of resonances (we can call them as FF-resonances) are the same thing 
when it comes to their mathematical essence \cite{fano, fano1, fesh}. The 
asymmetry is explained as a byproduct of the quantum-mechanical interference 
between two separate channels of the reaction. Namely, interference between 
a background (continuum of states) and a resonant (excitation of the discrete 
states) scattering process produces the asymmetric line-shapes. The two 
separate channels (closed and open) differ but they couple to each other. The 
resonance has an energy width that depends on the coupling between the channels.

In particle physics, such peaks are identified with unstable particles. 
Let us mention, however, that sometimes even the "ordinary" hadrons are
treated as FF-resonances \cite{rlj}. It is usually claimed that the symmetric 
shape is observed for resonances directly produced in particle collisions. 
Their characteristics are compiled by the PDG (Particle Data Group) 
\cite{china}. However, 
some asymmetry was recently experimentally noticed for narrow resonances (with 
c- and b-quarks). It was explained as a consequence of FF-effects 
\cite{len1, len2} with interference of bound states and continuum induced
by vacuum excitation of light quark-antiquark pairs leading to creation of 
D-mesons. 

The situation has strongly changed after the high energy nucleus-nucleus 
collisions became available. The created particles have to leak somehow from
the nuclear medium. Medium interactions may lead to some modification of their
characteristics. Really, there are numerous experimental data 
\cite{agak, adam, arna, damj, trnk, naru, muto, kozl, kotu, tser} about the
in-medium modification of widths and positions of prominent wide vector meson
resonances. Some of them even contradict each other. They are mainly obtained 
from the shapes of dilepton (decay products) mass and transverse momentum 
spectra in nucleus-nucleus collisions. Some excess over the expected 
symmetric shape of the distribution was observed. Dilepton spectroscopy 
directly probes the vector component of the spectral function of the hadronic 
medium. The dilepton mass spectra decrease approximately exponentially
with increase of masses but show peaks over this trend at some masses which
can be identified with prominent resonances. The $\rho $-meson peak is usually
the strongest one \cite{agak, adam, arna, damj} in the ratio 
$\rho :\omega :\phi =10:1:2$. Below, we concentrate on properties of
in-medium $\rho $-mesons with the special attention to be paid to the 
asymmetry of their shape.

Several approaches have been advocated for explanation of the observed excess
and properties of in-medium resonances \cite{pisa, hara, brow, bore, dusl, leup, 
rupp, chiu, elet, elio, mart, rapp, hees}. Most of them use effective hadronic 
Lagrangians to compute loop corrections and/or just hydrodynamics ideas
of the expanding fireball. See also the review \cite{haya} on cold nuclear 
matter effects. However, either positions, widths or heights presented some 
problems. Therefore we will not review these attempts in detail.
Here we concentrate not as much on their particular values which we just fit by the
corresponding parameters as on the asymmetry of the resonance shape.

\section{Experiment, Fano-Feshbach-effect and \\ Cherenkov gluons}

\begin{figure}[ht]
\includegraphics[width=\textwidth]{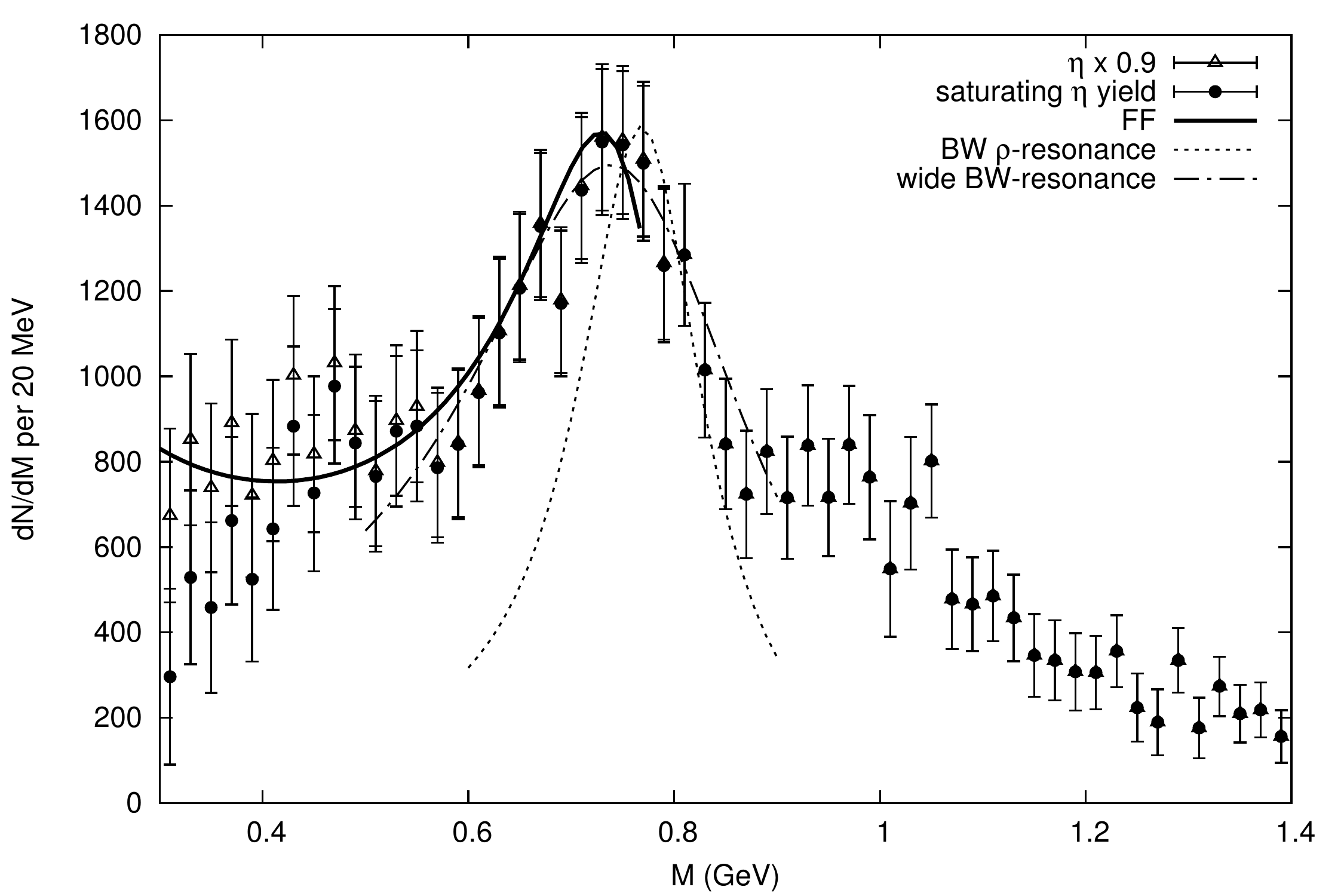}
 \caption{The spectrum of dileptons in semi-central collisions 
In(158 A GeV)-In measured by NA60-Collaboration \cite{damj} (points with error 
bars). The solid line shows that the fits by Eq. (\ref{sigma1}) with FF-effect
and by Eq. (\ref{ll}) with gluon Cherenkov effect taken into account coincide.
The dashed line corresponds to the Breit-Wigner shape of $\rho $-meson
with PDG-parameters \cite{china}. The dash-dotted line shows this shape 
with the modified width.}
\label{fitFF_PDG}
\end{figure}

The dilepton mass spectrum in semi-central In-In collisions at 158 AGeV
measured by NA60-Collaboration \cite{arna} is shown in Fig.\ref{fitFF_PDG}
by dots with error bars in the region of $\rho$ and $\phi , \; \omega$-mesons. 
Its asymmetry is easily seen with some excess in the low-mass wing. The shape 
is quite distinct from the familiar $\rho$-peak with PDG-parameters 
$m_{\rho}=775$ MeV, $\Gamma _{\rho}=149$ MeV \cite{china}
shown in Fig.1 by the dashed line.

The in-medium modification of $\rho$-meson parameters can not be accounted by
simple variation of them within the Breit-Wigner formula (\ref{bw}). It is
demonstrated by the dash-dotted line in Fig. 1 with fit parameters $m_r=775$ 
MeV, $\; \Gamma _r=336$ MeV (much larger width!) which does not reproduce the 
observed asymmetry. 

As described above, the interference of the continuum states and discrete 
levels of the reaction leads in quantum mechanics to the well known FF-effect 
\cite{fano, fano1, fesh}. They interfere with opposite phase on the two sides 
of the resonance as shown in \cite{fano1}. The resonance asymmetry is described 
by the following formula derived in Ref. \cite{fano1}:
\begin{equation}
\sigma = \frac {(q+\epsilon )^2}{1+\epsilon ^2}=1+\frac {q^2-1+2q\epsilon }
{\epsilon ^2 +1},
\label{sigma}
\end{equation}
where at the relativistic notation
\begin{equation}
\epsilon = \frac {M^2-m_r^2}{M\Gamma _r}.
\end{equation}
This expression replaces its non-relativistic form in the original paper 
\cite{fano} to reproduce properly the contribution due to
the Breit-Wigner resonance shape (\ref{bw}).
The parameter $q$ describes the relative strength of discrete states and
unperturbed continuum. The term linear in $\epsilon $ is in charge of asymmetry. 
After subtracting the constant background and normalizing the Breit - Wigner
part, one gets the following expression which can be fitted to experimental 
data of Fig.~\ref{fitFF_PDG} when multiplied by the overall normalization 
factor $C$:
\begin{equation}
C\frac {1-\sigma }{1-q^2} = C\frac {1-2q\epsilon /(1-q^2)}{\epsilon ^2 +1}.
\label{sigma1}
\end{equation}
Unfortunately, one can not use for the fitting procedure the experimental
data in the whole region of masses $M$ shown in Fig.~\ref{fitFF_PDG} because
the admixture of higher mass resonances is large in the right wing of the
Figure. Therefore we had to use the data in the smaller interval of masses
below the peak, i.e., in the left wing only.
The fit shown by the solid line in Fig.~\ref{fitFF_PDG} resulted in the 
value of $q\approx 0.363$ ($\Gamma_r=184$ MeV, $C=29$). Let us note that
the width is larger than the ordinary one. 

One concludes that the interference parameter $q$ is quite noticeable to
explain the shape of the resonance with excess of mesons in the 
left wing (smaller masses) of the resonance profile.

The admixture of the contribution of direct states to this effect compared
to the influence of the continuum was estimated in \cite{fano1} equal
$\pi q^2/2$ which amounts to about 0.21 in our case. Thus we conclude that
the interference of the continuum with quasibound states is quite important.

The above treatment is based on general quantum-mechanical principles and does
not reveal what particular mechanisms are in charge of the interfering open and 
closed channels. One can propose emission of Cherenkov gluons in the nuclear 
media \cite{drem} as one of these channels that explains a possible source of 
the left-wing asymmetry. 

The necessary condition for Cherenkov effects to be observable within some 
energy interval is an excess of the refractivity index of the medium $n$ over 1. 
It is well known for ordinary media that such an excess happens 
due to electromagnetic interactions just in
the left wing of any resonance (e.g., see Fig. 31.5 in \cite{fls}).
According to general formulas (e.g., see \cite{akpo}) this excess is 
proportional to the real part of the forward (depicted by 0 below) 
scattering amplitude. 
\begin{equation}
\Delta n ={\rm Re }n -1 \propto {\rm Re }F(M,0)>0. 
\label{delta}
\end{equation}
For the nuclear quark-gluon medium this requirement should be fulfilled for
the chromopermittivity of gluons \cite{drle}. The real part of the
Breit-Wigner amplitude leading to Eq. (\ref{bw}) is positive just within the 
low-energy (left) wing of any resonance described by this equation (see, e.g., 
\cite{elet, elio}). Herefrom one gets the general prediction that the shape of 
{\it any} resonance formed in high energy nuclei collisions must become 
asymmetric with some excess within its left wing compared to the usual 
Breit-Wigner shape. One could expect that some collective excitations of the 
quark-gluon medium may contribute in these energy intervals in addition to the 
traditional effects. Since the probability of Cherenkov radiation is 
proportional to $\Delta n$ the asymmetry must be proportional to it. Then the 
dilepton mass distribution must get the shape (the formula in \cite{drem} is 
slightly corrected):
\begin{equation}
\frac{dN_{ll}}{dM}=\frac {A}{(m_{r }^2-M^2)^2+M^2\Gamma _r^2}
\left(1+w_r\frac{m_{r}^2-M^2}{M\Gamma _r}\Theta (m_{r}-M)\right).  \label{ll}
\end{equation}
This formula was also used to fit experimental data in Fig.~\ref{fitFF_PDG}.
The second term is due to the coherent Cherenkov gluon response of the medium
to the penetrating quark 
proportional to the real part of the amplitude. It is in charge of the
observed asymmetry. It vanishes at energies above the resonance peak $M>m_r$
because only positive $\Delta n$ lead to Cherenkov effects. Here, we take into 
account that the ratio of real to imaginary parts of Breit-Wigner amplitudes is
\begin{equation}
\frac {{\rm Re}F(M,0)}{{\rm Im}F(M,0)}=\frac {m_r^2-M^2}{M\Gamma _r}.
\label{ratio}
\end{equation}  
The weight of the second term is described by the only adjustable parameter 
$w_r$ for a given resonance $r$. As we see, the general structure of 
Eqs (\ref{sigma1}), (\ref{ll}) is the same with 
\begin{equation}
w_r=2q/(1-q^2).
\label{wq}
\end{equation}
The adjusted parameters were obtained from the independent fit to 
experimental points in Fig. 1. 
They are $A=25, \; \Gamma _r=184$ MeV,$\; w_r=0.838$. The relation (\ref{wq}) 
is well fulfilled. Therefore both fits according to (\ref{sigma1}) and 
(\ref{ll}) practically coincide and are shown in Fig. 1 by a single solid line.

The quantum interference of continuum and quasibound states is at the origin 
both of asymmetric resonances and of the classical phenomenological prescription 
of $\Delta n>0$, which is required for Cherenkov effect. Therefore, FF-effect 
can serve as the quantum-mechanics foundation of classical Cherenkov effect 
in general. The overlap of both fits demonstrated by the solid line 
in Fig.~\ref{fitFF_PDG} supports this conclusion.

\section{QCD perspective}

At that stage one is tempted to speculate about the QCD interpretation of such 
a statement. High energy nuclei collisions give rise to a state of the boiling 
quark-gluon matter (plasma?). The numerous quark-antiquark pairs with different 
colors and masses (energy in the center of mass system of the pair) are 
produced there. The color-neutral pairs whose mass fits the Breit-Wigner 
shape form the resonance. 

\begin{figure}[ht]
\centerline{\includegraphics[width=0.5\textwidth]{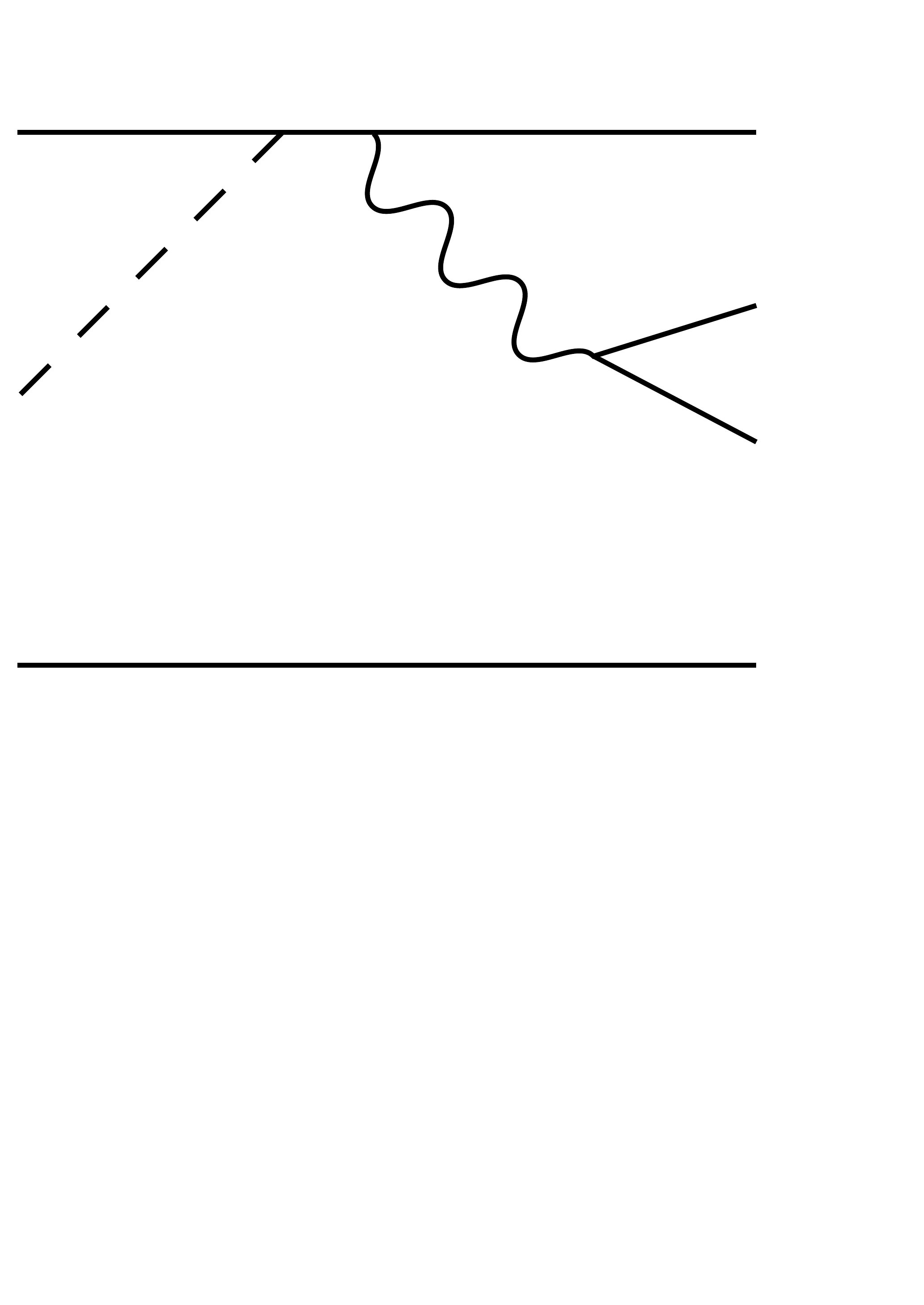}}
 \caption{Origin of the tetraquark state.
One of the quarks of the initially colored pair interacts within the medium
(shown by the dashed line attached to it),
changes its color and gets excited, finds its partner for the color-free bound 
state within the low-mass wing of a resonance, gets de-excited by emission of 
Cherenkov gluon that leads to the tetraquark states in this wing.}
 \label{diagram}
\end{figure}

However, most pairs in the plasma are in a color-octet state and may not create 
resonances. They are considered as a continuum. After interaction with a gluon
or collective excitation in the medium (shown by the dashed line in Fig. 2)
some quarks can change the color and 
get excited. If such a quark finds a partner to form a color-neutral pair in the 
left wing of the resonance it can emit Cherenkov gluon as allowed by the 
chromopermittivity (argued above in classical terms). This gluon transforms to
the quark-antiquark pair. That is how the color-neutral ${\bar q}qg$-component 
(or four-quark component) of the resonance can be formed from the initial 
colored two-quark state. Thus beside the common color-neutral 
${\bar q}q$-component in the low-mass branch of the $\rho$-meson peak there 
appears new ${\bar q}qg$-component (as well as others, possibly). Using the 
wave-line notation for the Cherenkov gluon shown in Fig.~\ref{diagram} one 
gets a tetraquark state which contributes to the left-wing cross section.

The distributions of the decay products of the left-wing and right-wing states 
of asymmetric heavy resonances created in the process of nucleus-nucleus 
collisions can slightly differ. If tetraquarks have additional decay modes 
compared to dimers, these modes can serve for identification of tetraquarks. 
Then these special modes should reveal themselves only in the low-mass wing of 
an asymmetric resonance. It is a distinctive feature of ${\bar q}qg$-Cherenkov 
states. That is more probable for heavy resonances because the presence of a 
heavy quark in heavy flavor hadrons provides an additional variety of decay 
channels with new energy scales. Even though the admixture of new decay 
channels is small, their search deserves special attention.

Surely, the emission of Cherenkov gluons asks for interaction of "$\rho $-meson 
quarks" within the medium (gluon, collective modes...?). Such an 
interaction can result either in continuum or quasistable states. In our case, 
the nature of the primary interaction with $\rho $-meson quarks initiating
emission of Cherenkov gluons is however left unknown yet. The main conclusion 
is that some nuclear forces initiate sometimes the additional binding 
to tetraquarks by emitted Cherenkov gluons of two otherwise independent 
unbound color-octet quarks, created in the nuclear medium. 

The formation of the weakly bound triple-states (${\bar q}qg$) in the 
collisions of three particles when two-particle forces are too weak to produce 
bound dimers is known as Efimov effect \cite{efim, ferl}. In all the cases
the role of the third component is crucial for experimental observation of
this effect. It is important to reveal the physics nature of the component.
In simplest models, it was ascribed to light quark pairs produced in vacuum 
in case of narrow resonances \cite{len1, len2} and to Cherenkov gluons for 
wide resonances \cite{drem}.  

\section{Conclusion}

To conclude, we have shown that asymmetry of vector mesons produced in 
nuclear collisions can be successfully described as Fano-Feshbach effect
and further interpreted in terms of emission of Cherenkov gluons as a
particular detalization of the quantum interference pattern.

From the theoretical side, models of (collective?) excitations in the nuclear
medium which help to get an insight to this problem are welcome.

From the experimental side, the error bars in experiments with $\rho $-mesons 
are quite large. There are some plans to improve experimental accuracy up to
two orders of magnitude (private communication). Very little is still known 
about other resonances but the low-mass asymmetry seems universal and gives 
some hope for further progress. 

The dilepton spectra and, especially, the asymmetry of vector mesons produced 
in nuclei collisions deserve further precise experimental studies at 
RHIC and LHC (see discussion in \cite{rapp1}). 

\medskip

{\bf Acknowledgments}

\medskip 
 
We are grateful for support by the RFBR-grant
14-02-00099  and the RAS-CERN program.

\end{document}